# Optical phase modulation by natural eye movements: application to time-domain FF-OCT image retrieval

VIACHESLAV MAZLIN,[1,*] PENG XIAO,[1,2] KRISTINA IRSCH,[3,4] JULES SCHOLLER, [1,5] KASSANDRA GROUX,[1] KATE GRIEVE,[3,4] MATHIAS FINK[1] AND A. CLAUDE BOCCARA[1]

[1]*ESPCI Paris, PSL University, CNRS, Langevin Institute, 1 Rue Jussieu, 75005 Paris, France*
[2]*State Key Laboratory of Ophthalmology, Zhongshan Ophthalmic Center, Sun Yat-sen University, 510060, Guangzhou, China*
[3]*Vision Institute, Sorbonne University, CNRS, INSERM, 17 Rue Moreau, 75012 Paris, France*
[4]*Quinze-Vingts National Ophthalmology Hospital, 28 Rue de Charenton, 75012 Paris, France*
[5]*Wyss Center for Bio and Neuroengineering, Chem. des Mines 9, 1202 Geneva, Switzerland*
*mazlin.slava@gmail.com

**Abstract:** Eye movements are commonly seen as an obstacle to high-resolution ophthalmic imaging. In this context we study the natural axial movements of the in vivo human eye and show that they can be used to modulate the optical phase and retrieve tomographic images via time-domain full-field optical coherence tomography (TD-FF-OCT). This approach opens a path to a simplified ophthalmic TD-FF-OCT device, operating without the usual piezo motor-camera synchronization. The device demonstrates in vivo human corneal images under the different image retrieval schemes (2-phase and 4-phase) and different exposure times (3.5 ms, 10 ms, 20 ms). Data on eye movements, acquired with a 100 kHz spectral-domain OCT with axial eye tracking, are used to study the influence of ocular motion on the probability of capturing high-signal tomographic images without phase washout. The optimal combinations of camera acquisition speed and amplitude of piezo modulation are proposed and discussed.

## 1. Introduction

The human eye is constantly moving. Many of the movements are caused by the unconscious physiological processes such as heartbeat, breathing and brain-induced saccades [1,2], and therefore cannot be completely suppressed. Ophthalmic imaging instruments produce an image by collecting light over a certain amount of time (e.g. integration time of the detector). When this time spans several milliseconds, the eye cannot be considered static, the movements affect the image quality and should be considered.

Optical coherence tomography (OCT) [3,4] is the clinical gold standard method for cross-sectional imaging of the eye. The principle of OCT is based on broadband light interferometry, which allows one to discriminate different layers of the imaged sample. Interferometry is susceptible to sample motion, in particular motion manifesting itself in the axial direction, which can induce optical phase washout and phase wrapping phenomena [5,6]. In this article we will refer to phase washout as to a process of tomographic signal fading due to additional optical phase shift, induced by the axial sample motion during the acquisition of the elementary tomographic data. In the example of the conventional OCT, the tomographic signal completely fades, if the axial movement of the sample produces a $\pi$ optical phase shift (corresponding to a movement of a few hundred nanometers) during the acquisition of a 1D axial reflectivity profile (A-line) of the sample. Phase wrapping happens, when the sample motion produces a large phase-shift lying outside of the [-$\pi$, $\pi$] interval. In the conventional OCT the latter does not necessary result in signal loss, but creates ambiguities in the measurements. The degree, to which the tomographic signal is affected by the movements depends on the particular type of OCT.

Today spectral-domain OCT (SD-OCT) is the most commonly used type of OCT. SD-OCT reconstructs 2D cross-sectional image (B-scan) by stacking scanned A-lines, with each A-line being acquired using a line-scan camera embedded into a spectrometer. Line-scan cameras have a typical integration time of a few microseconds – sufficiently short to avoid phase washout during the A-line capture. If necessary, the residual A-line artifacts can be completely suppressed by using pulsed light, that effectively shortens the exposure time of the camera [7]. Motion artifacts are however more pronounced in the final B-scan image, as it is formed by point-by-point laser beam scanning in the en face plane. These artifacts have become much less visible in recent years given the remarkable progress in increasing the imaging speed, today reaching over 250 000 A-scans/s [8,9].

Swept-source OCT (SS-OCT) [10,11] is another modality, which acquires spectral data like SD-OCT, but employs a tunable wavelength laser source (swept-source) and a photodetector instead of spectrometer. Although the photodetector needs to acquire data sequentially over a period of time, this period can be very short thanks to the fast A/D acquisition electronics operating at GHz speeds. It takes less than a μs to capture the entire spectrum, which diminishes the phase washout effect [7,12]. Moreover, the image acquisition rate can be increased to several MHz [13,14], leading to further suppressed motion artifacts in the final B-scan and volumetric images.

While the above OCT methods have a remarkably fast imaging speed along the axial direction (~ μs), imaging in the lateral direction takes a longer time (~ ms), as it typically requires a point-by-point laser beam scanning in a 2D plane with a galvanometer mirror system. In an alternative to point-scanning, a new class of parallel acquisition methods were proposed, including line-field SD-OCT (LF-SD-OCT) [15] and Fourier-domain full-field OCT (FD-FF-OCT) (also known as FF-SS-OCT) [16]. These instruments increased dimensionality of data acquisition by substituting the 0D photodetectors and 1D line-scan cameras with 2D cameras. This enabled increase of the effective A-line rate up to ~ 2 MHz for LF-SD-OCT [17,18] and ~ 40 MHz for FD-FF-OCT [19–21]. Similar to SS-OCT, FD-FF-OCT needs to acquire spectral information over time. It takes typically ~ 10 ms to capture 500 images at a 60,000 images/s frame rate and retrieve the entire spectrum. This is a relatively long time (~ 1,000 times longer than the SS-OCT exposure time), during which the eye can potentially shift by more than several hundred nanometers, leading to $\pi$ phase shift and phase washout. In practice, the washout effect is less pronounced, potentially because the spectral information is captured sequentially in small spectral samples, each of them being acquired in a fraction of the total exposure time (~ 10/500 = 20 μs) [12]. The remaining eye movement artifacts can be suppressed via numerical motion correction in post-processing [22]. LF-SD-OCT acquires the spectrum in a single camera shot, like SD-OCT, which in principle makes it less sensitive to phase washout, assuming the camera exposure time is sufficiently short.

Another type of parallel OCT is time-domain full-field OCT (TD-FF-OCT). Contrary to the spectral-domain OCT methods, which obtain the best results with the static interferometer arms (e.g. no optical phase shift), TD-FF-OCT by design requires phase shifting between the arms. This can be achieved by modulating the optical path length of the reference arm with a piezo-motor [23], by using the two physically separated synchronized cameras [24,25], by tilting the reference beam axis in holographic approach [26], or by using the metabolic cell dynamics [27,28], etc. Recent developments in fast and high-full-capacity (FWC) 2D cameras enabled eye imaging applications of TD-FF-OCT, such as 600 MHz (en face pixels/s) angiography in the anterior human eye with capability of tracking individual blood cells [29], corneal [29–31] and retinal [32,33] imaging. In most cases the tomographic images were reconstructed from 2 phase-modulated (with piezo-motor) camera frames captured in ~ 4 ms.

In this article we study the effect that the eye movements play during the acquisition of tomographic images in TD-FF-OCT. We show that the natural ocular movements can (with high probability) shift the eye of the fixated subject axially by many hundreds of nanometers during a millisecond. This suggests that the eye movements themselves can be used for modulation of the optical phase, substituting the piezo-modulation. In this way, eye movements are seen not as an obstacle, contrary to the methods above, but become an essential part of the image reconstruction process. We search for the optimal camera speed and amplitude of piezo modulation that would enable the highest signal TD-FF-OCT images in presence of eye movements.

## 2. Materials and Methods

### 2.1 SD-OCT with axial eye tracking

We measured the axial movements of the eye (in human cornea in vivo) using a 100 kHz SD-OCT device, resembling that in article [29]. In brief, the SD-OCT is based on a commercial GAN510 (Thorlabs, USA) (Fig. 1).

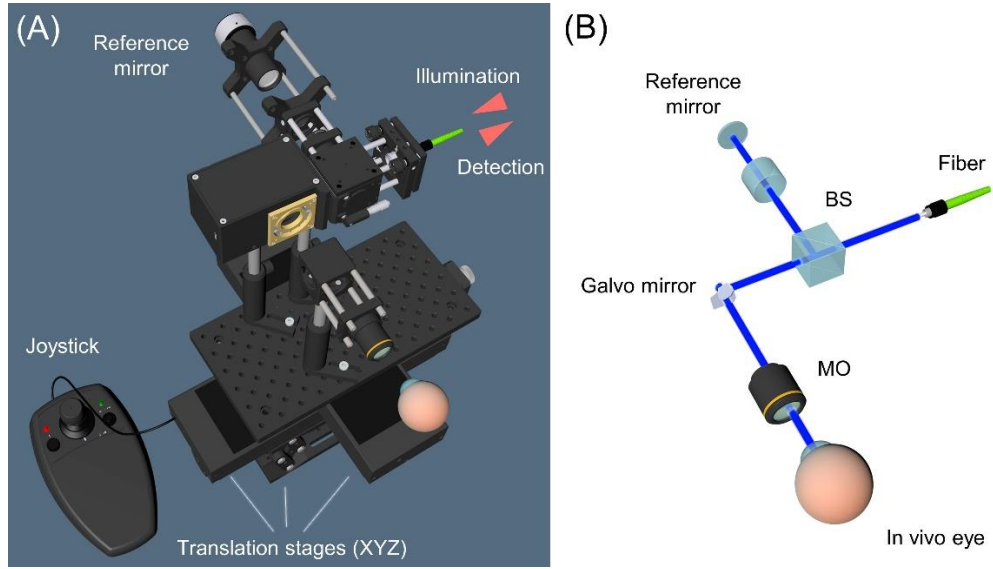

**Fig. 1.** Detailed (A) and schematic (B) illustrations of SD-OCT, used to measure the axial eye movements. BS: beamsplitter. MO: microscope objective. Spectrometer is not shown. Interferometer is mounted on XYZ translation stages, useful for aligning with respect to the eye.

It consists of the interferometer with 10× microscope objectives (LMPLN10XIR, Olympus), detecting spectrometer, galvanometric mirror system (OCTP-900(/M), Thorlabs, USA) and superluminescent diode (SLD). The galvanometric mirror rapidly scans a collimated light beam laterally at 100 kHz to form a 2D cross-sectional B-scan image. The SLD has central wavelength of 930 nm and spectral bandwidth of 40 nm. Dimensions of the B-scan were 1.25 mm (lateral) × 2.7 mm (axial).

We performed axial tracking by detecting the position of the reflective peak signal from the air-tear film interface. As this is a single non-overlapping peak, its position can be detected with a better resolution than 6.9 μm (in the cornea) as derived with the Rayleigh criterion and the spectral bandwidth. The peak was located via Labview software (National Instruments, USA) by averaging the B-scan along the lateral dimension and applying the Peak detector virtual instrument (National Instruments, USA). Axial movements of the eye were tracked at 180 Hz, locating a new corneal surface position every 5.5 ms, during 20 – 30 seconds.

### 2.2 TD-FF-OCT with enabled and disabled piezo motor

We used TD-FF-OCT to obtain en face tomographic images of the human cornea. The design of TD-FF-OCT is similar to that reported in [29–31]. In brief, the device is a Linnik interferometer, equipped with the two microscope objectives, a 2D camera and a light source of low spatial and temporal coherence, precisely light emitting diode (LED) (Fig. 2).

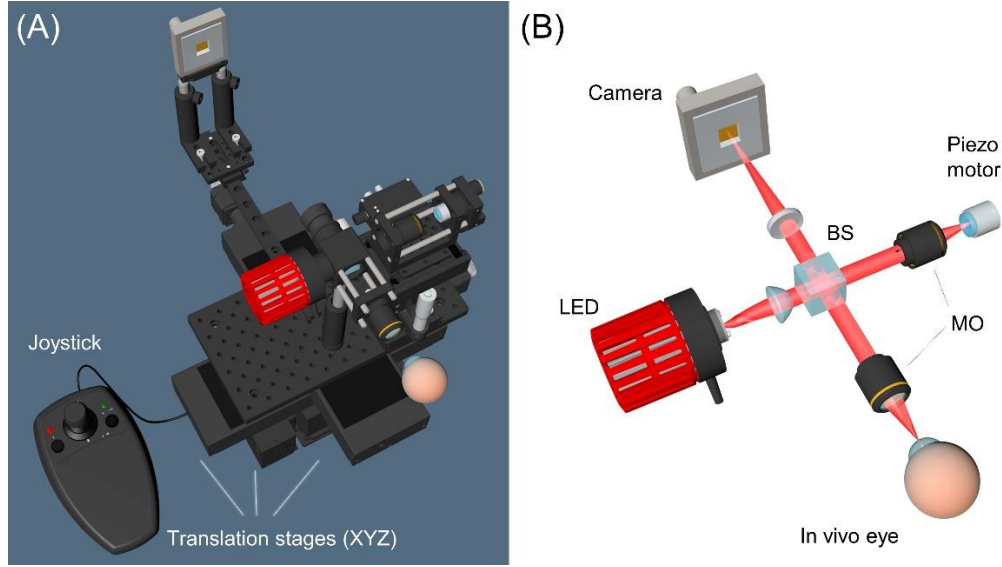

**Fig. 2.** Detailed (A) and schematic (B) illustrations of TD-FF-OCT, used to image the in vivo human cornea. BS: 50:50 beamsplitter. MO: microscope objectives. Interferometer is mounted on XYZ translation stages, useful for aligning with respect to the eye. The piezo motor in the reference arm, that holds the mirror, may be disabled, leaving the axial movements of the eye to be responsible for optical phase modulation.

Microscope objectives (LMPLN10XIR, Olympus, Japan) and a 250 mm tube lens provide a 13× magnification (1.1 mm field of view), 18 mm working distance and 1.7 μm lateral resolution. The LED (M850LP1, Thorlabs, USA) has a central wavelength of 850 nm and a spectral bandwidth of 30 nm, resulting in an axial resolution of 7.7 μm (in the cornea). The sensor is a 1440 × 1440 pixels CMOS camera with a high 2Me$^-$ full well capacity (Q-2A750-CXP, Adimec, Netherlands). Camera imaging speed was 550 frames per second (fps) (1.75 ms camera exposure time). An absorptive neutral density (ND) filter of 4% reflectivity (NENIR550B, Thorlabs, USA) was used as a reference mirror to achieve high detection sensitivity and avoid multiple reflections. The filter was mounted on a piezo motor (STr-25/150/6, Piezomechanik GmbH, Germany), synchronized with the camera, to capture several phase-shifted consecutive images. Modulation amplitude of the piezo motor could be adjusted or disabled to perform phase-shifting only with the axial movements of the eye. The interferometer was mounted on the three high-load translation stages (two NRT150/M, Thorlabs, USA and MLJ150/M, Thorlabs, USA) enabling positioning in three axes.

*2.3 In vivo imaging*

Approval for the study was obtained (study number 2019-A00942-55), in conformity with French regulations, from the CPP (Comité de Protection de Personnes) Sud-Est III de Bron and ANSM (Agence Nationale de Sécurité du Médicament et des Produits de Santé). The study was carried out on two healthy subjects, which was confirmed with eye examinations at the Quinze Vingts National Ophthalmology hospital preceding the experiment. Prior to experimental procedures, which adhered to the tenets of the Declaration of Helsinki, informed consent was obtained from the subjects after the nature of the study was explained. The subjects were asked to rest the chin and temple on a standard clinical headrest and look at a fixation target within the device. Examination was non-contact and did not involve instillation of eye drops into the eye. The SD-OCT used a scanning light beam and the TD-FF-OCT used a pulsating light to keep the light exposure significantly below the maximum permissible level, as detailed in [29].

## 3. Results

### 3.1 Analysis of axial eye movements

Axial eye movements were measured using a fast SD-OCT with corneal tracking (Fig. 1). A new corneal position was recorded every 5.5 ms over a duration of 30 seconds (Fig. 3). Such a long period enables exploration of wide diversity of eye movements. The amplitude of movements ranged from -150 μm to 150 μm. The main contributors to the motion were fast heartbeat around 1 Hz with higher harmonics at 2 Hz, 3 Hz, etc. and slow breathing at 0.34 Hz, in agreement with the literature [1].

Our primary goal is to study the movements on a shorter (millisecond) time scale. On that scale eye position changes smoothly with time (see zoom in Fig.3), allowing a linear interpolation of the experimental data with a finer time step than experimentally recorded 5.5 ms. In order to analyze the influence of eye movements, we divided the entire timeline into the sequences of virtual camera frames (with exposure times of 0.1 ms, 1 ms, etc.) and checked how the corneal position and optical phase change from one virtual frame to another.

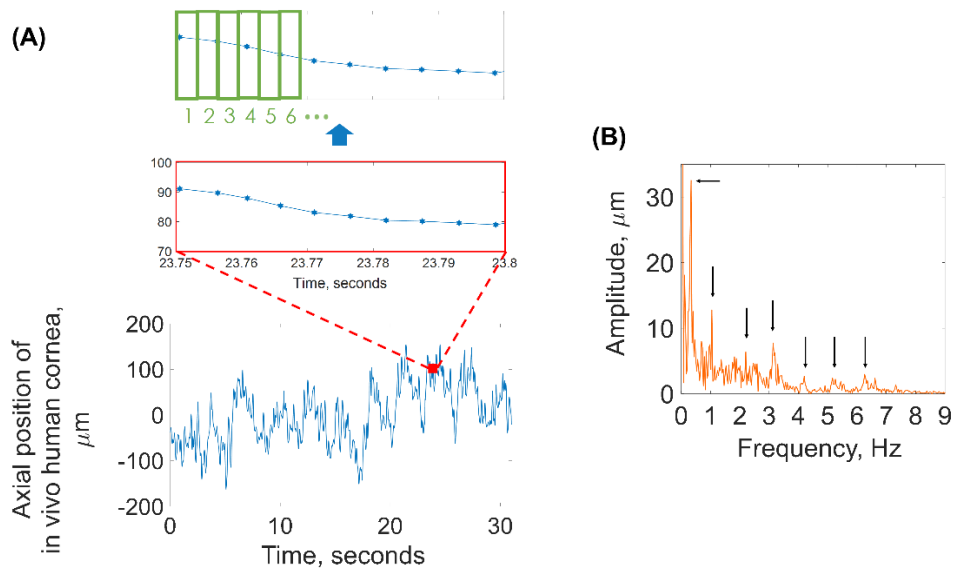

**Fig. 3.** (A) Axial movements of in vivo human cornea, measured with SD-OCT. The source data underlying the plot are provided in **Data File 1**. The long period of recording enables exploration of wide diversity of eye movements. Red square (zoomed image) highlights the linearity of eye movements on a ms time scale. The entire plot can be divided into the imaginary time windows (e.g. 1 ms, 3.5 ms, 10 ms etc.), shown in greem. (B) Fourier transform of the main

plot (A). Peaks corresponding to breathing at 0.34 Hz and heartbeat at 1 Hz, 2 Hz, 3 Hz are visible in agreements with the literature [1].

Fig. 4 shows the distributions of all possible axial corneal shifts, optical phase shifts and velocities happening between the different consecutive frames over the entire 30 seconds timeline. As expected, the shorter the time of the frame (exposure time), the smaller is the eye displacement happening during this time, and therefore, the smaller is the influence of the movements on the acquired images. During the exposure times below 0.1 ms the eye shifts by less than 100 nm. This increases the optical path length of light by 200 nm (taking into account the double optical pass to the sample and back), equivalent to an optical phase shift of $\pi/2$. Here we considered a light wavelength of 850 nm, commonly used in ophthalmology. This highlights that instruments with even shorter (microsecond) exposure times (SD-OCT, SS-OCT, LF-SD-OCT) are almost insensitive to fringe washout ($\pi$ phase shift). On the contrary, during exposure times longer than 1 ms the eye can shift by 1 μm and more, leading to large optical phase shifts of $\pi$ and more. Insteresting to note that, while such large movements can potentially lead to phase washout in spectral-domain methods, time-domain techniques (e.g. TD-FF-OCT) can potentially use them to their advantage for optical phase modulation. Fig. 4(D) highlights that large phase shifts ($> \pi/2$) account for almost 80% of all shifts within 1 ms, for almost 100% within 10 ms and for only 3% within 0.1 ms and below.

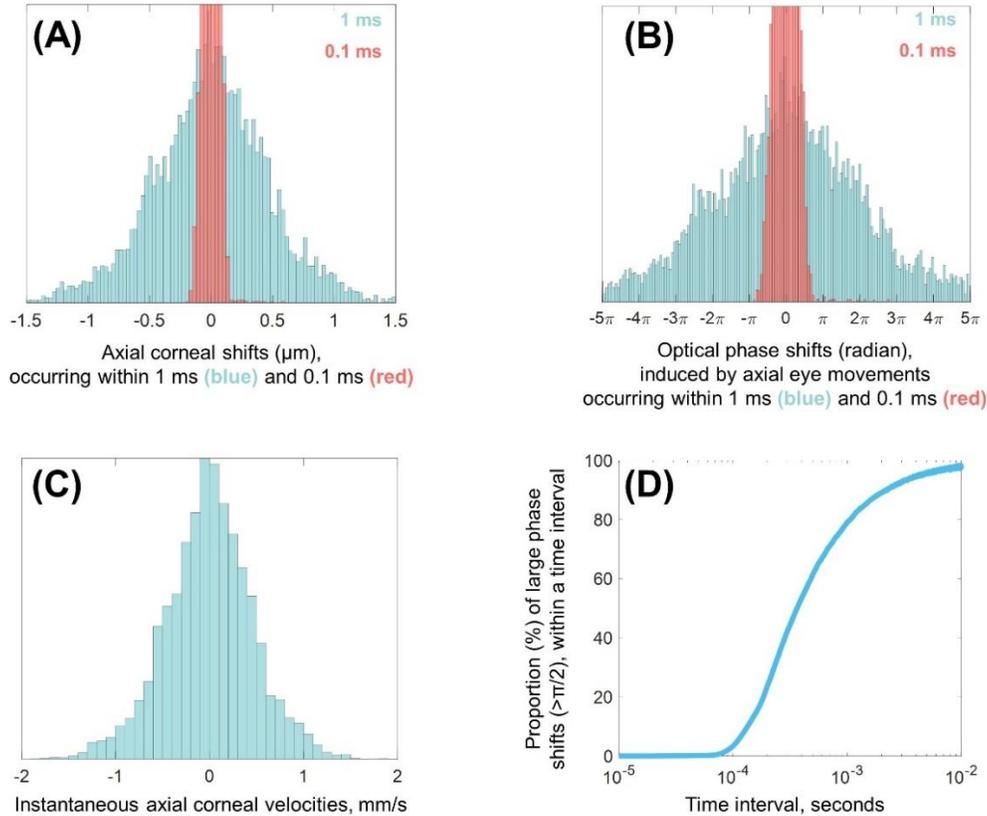

**Fig. 4.** (A) Distributions of axial corneal position shifts during different time intervals. Data is obtained from Fig. 3(A), divided into imaginary time frames. (B) Distribution of optical phase shifts calculated from (A), assuming an 850 nm light wavelength. (C) Distribution of instantaneous axial corneal velocities. (D) Proportion of a large phase shift ($> \pi/2$) among all shifts happening during different time intervals. The graph is calculated from (B). It should be noted that large phase shifts ($> 2\pi$) also can produce a high tomographic signal, but for some OCT methods this may require additional phase unwrapping.

*3.2 Analysis of phase modulation with eye movements in TD-FF-OCT*

Below we will analyze the influence of eye movements on TD-FF-OCT signal. The simplest TD-FF-OCT image retrieval scheme requires two consecutive camera frames to reconstruct a tomographic image [34]. Below by the exposure time we will mean the time, required to capture the tomographic TD-FF-OCT image (equal to the time of 2 camera frames). The tomographic signal is the highest when the optical phase shift (supposed here to be instantaneous) between those frames equals $\pi$, $3\pi$, etc. and is absent at 0, $2\pi$, $4\pi$, etc. The higher order phase shifts of $5\pi$, $7\pi$ etc. can also produce the highest tomographic signal, as long as the axial corneal shift stays within the coherence length. The latter is typically true for ophthalmic TD-FF-OCT [30], given that during its exposure time of 3.5 ms (2 camera frames, each frame 1.75 ms), the axial corneal shifts are still within its coherence gate thickness (axial resolution) of 7.7 μm (Fig.4(A)).

The intensity on the camera can be written generally, as:

$$I = \frac{I_0}{4} \cdot \left\{ R_{inc} + R_{coh-sam} + R_{ref} + 2 \cdot \sqrt{R_{coh-sam} \cdot R_{ref}} \cdot \int_{T_0}^{T_1} \cos(\varphi + \phi(t) + \psi(t)) dt \right\}, \quad (1)$$

where $I_0$ is the photon flux of illumination, $R_{inc}$ is the 'incoherent' reflectivity from all the sample structures and optical system parts outside of the coherence volume, $R_{coh-sam}$ is the 'coherent' reflectivity from all sample structures lying within the coherence volume, $R_{ref}$ is the reflectivity from the reference mirror, $\varphi$ is the phase difference between the sample and reference signals (denoting fringes in the 2D image), $\phi(t)$ is the additional phase, modulated with the step function in time by the piezo motor, $\psi(t)$ is the additional phase, modulated in time by the movements of the eye and $T_1 - T_0$ is the camera exposure time. Note that the locations of the pixel in the 2D image $(x, y)$ are omitted from the arguments.

For simplicity we select consider one pixel $(x, y)$ on the camera, for which the spatial phase $\varphi(x, y) = 0$. In order to remove the incoherence part and retrieve the interference amplitude, two images are subtracted and the modulus is taken:

$$|I_1 - I_2| = const \cdot \left| \int_{T_0}^{T_1} \cos(\phi(t) + \psi(t)) dt - \int_{T_1}^{T_2} \cos(\phi(t) + \psi(t)) dt \right|, \quad (2)$$

where $const = I_0^2 \cdot R_{coh-sam} \cdot R_{ref}$, while $T_1 - T_0$ and $T_2 - T_1$ are the exposure times for the first and second frames respectively. The module term determines the strength of tomographic signal. If the optical phase shift between the consecutive camera frames is small, then the term in the modulus is small and the tomographic signal is low.

For calculations we will consider the typical parameters for in vivo TD-FF-OCT [28]: 850 nm light wavelength and exposure time of 3.5 ms (2 camera frames, each frame 1.75 ms). We will also rely on a common TD-FF-OCT piezo-modulation scheme, for which $\phi(t)$ equals to 0 every first frame and to $\pi$ every second frame.

The distribution of possible eye motion induced phase shifts $\psi(t)$ was computed using 3.5 ms exposure time and the axial movements from Fig. 3. By inserting this distribution into Eq. 2, we obtain the distribution of possible TF-FF-OCT signals in the presence of eye movements (Fig. 5). Signal is normalized to 1.

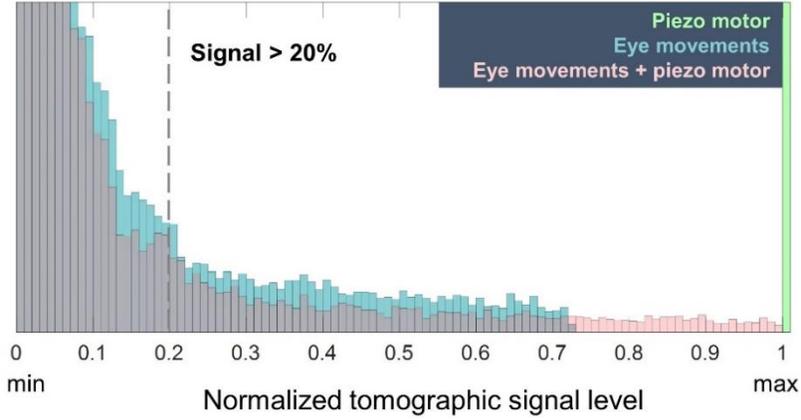

**Fig. 5.** Distribution of possible TD-FF-OCT signals (normalized) under different sources of phase modulation. The histograms are calculated from Eq. 2 considering the experimental data in Fig. 3, 3.5 ms exposure time (2 camera frames, each frame 1.75 ms) and 850 nm light wavelength. The signal is the highest (equal to 1), when only the piezo motor contributes to phase modulation in the absence of eye movement. In case of phase modulation with axial eye movements, the tomographic signal can take only values below 0.7. As discussed below, this results from integration of continuous function of time (phase change with eye movements is a continuous function of time). The addition of discontinuity, such as made with a fast step modulation with a piezo motor, can create tomographic signals of any value. 20% denotes the arbitrary selected threshold of 'sufficient' tomographic signal that is used for calculating signal statistics.

As expected in case of a static sample and $\pi$ piezo motor modulation between the consecutive frames, TD-FF-OCT signal is always at maximum. A rather unusual distribution of values is obtained if we consider modulation by the in vivo movements of the eye only (without piezo motor). More precisely, if we rely only on the phase-shifting by eye movements, we cannot obtain a TD-FF-OCT signal higher than 70% of the theoretical maximum. Interestingly, this cannot be explained by the movements of the eye being too small, because, as we have seen above, they frequently give the $\pi$, $2\pi$ and higher phase-shifts between the consecutive frames (for 3.5 ms exposure time). The reason for such behavior is that the axial eye position (and optical phase) is a continuous function of time. More precisely, in Fig. 3 we saw that on a short ms time scale, axial eye position (and optical phase) is a linear function of time $\psi(t) = a \cdot t$. Then, substituting in Eq. 2 and assuming no piezo modulation, one obtains:

$$|I_1 - I_2| = const \cdot \frac{|\sin(aT_1) - \sin(aT_0) - \sin(aT_2) + \sin(aT_1)|}{a} \quad (3)$$

For simplicity we can arbitrarily select the start of the timeline $T_0 = 0$, $T_1 = T$, $T_2 = 2T$. Then, $a = (\psi(T) - \psi(0))/T$ and

$$|I_1 - I_2| = const \cdot \left| \frac{\sin(\tau) \cdot (1 - \cos(\tau))}{\tau} \right|, \quad (4)$$

where $\tau = \psi(T) - \psi(0)$. This normalized function can have values only below 0.7 (Fig. 6). This means that TD-FF-OCT signal, modulated by the eye movements (which change linearly on a ms time scale), is limited to 70% of the theoretical maximum.

On the contrary, if in addition to phase shifting with eye movements, we add a step-like discontinuous phase shifting with piezo motor, Eq. 2 becomes:

$$|I_1 - I_2| = const \cdot \left| \int_{T_0}^{T_1} \cos \psi(t) \, dt - \int_{T_1}^{T_2} \cos\left(\pi + \psi(t)\right) dt \right| \quad (5)$$

$$|I_1 - I_2| = const \cdot \left| \frac{\sin(\tau)}{\tau} \right| \quad (6)$$

And the function can take any value, which confirms the results in Fig. 5.

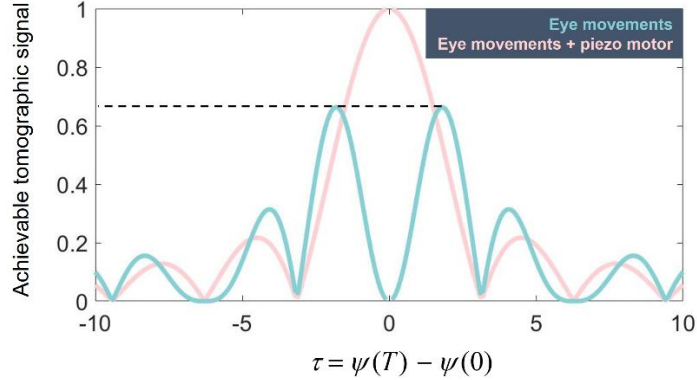

**Fig. 6.** Values that tomographic signal can take in the cases of modulation with eye movements and piezo motor or with eye movements only.

It is interesting to note that the lowest tomographic signal is achieved if the eye movements produce π phase shift during the acquisition of a single camera frame. This is explained by the fact that as the eye moves continuously and linearly at millisecond timescale, camera integrates all the phases between 0 and π, which leads to phase washout. In the absence of piezo modulation, the highest tomographic signal is reached, when the eye movements produce a phase shift close to π/2.

As we have seen above, the tomographic signal can take a variety of values depending on the eye movements at each time moment. What follows is that the discussion about the deterministic signal should be replaced with the discussion about the signal statistics. Two quantities are useful, when evaluating the signal statistics: 1) an average signal level over time and 2) probability of obtaining a signal above a certain threshold level. The latter measure can be used to find the probability of getting images with sufficient tomographic signal or equivalently the proportion of images with tomographic signals among all images. The threshold for selecting 'sufficient' signal is arbitrary and we chose it at 20% from the theoretical maximal signal. Main conclusions below will stay true for the other threshold levels.

Fig. 7 shows how these two quantities change depending on the exposure time and phase shift of piezo modulation. When the exposure time is above 10 ms, piezo movement has no influence on the probability of obtaining tomographic images and the phase modulation is dominated by the ocular movements. Proportion of the images is low at about 10%. Reduction in the exposure time leads to increase in the chances of getting tomographic images: 3.5 ms exposure – 20%, 1 ms exposure – 50% (every second image). Only below 1 ms exposure the piezo modulation begins to grow its contribution into the total phase modulation. Below 0.1 ms the piezo modulation becomes essential: with piezo shift of π – 85% tomographic images, without piezo shift – 10%. As an intermediate conclusion: if one is interested to rely only on the eye movements for phase modulation, it is worth to keep the exposure time at 1 ms or higher, for which the ocular motion is still sufficiently fast to provide the required phase shift and for which the proportion of tomographic images is the highest (50%). The additional phase modulation with the piezo motor does not play significant role at the exposures above 1 ms, however at shorter exposures it becomes a requirement, because the eye movements are too slow to provide a sufficient phase shift. 100% of images can be tomographic with piezo modulation and very short times below 0.1 ms.

The average signal level in tomographic images also changes with different exposure times and with different phases of piezo modulation. At a long exposure time of 10 ms the average signal is 40% and independent from the piezo motor. Decreasing the exposure time has similar effect on average signal level as on the proportion of tomographic images: piezo modulation starts to directly influence the signal level and the average signal becomes maximal for π phase shift. At 0.1 ms average signal reaches 70% at the maximum.

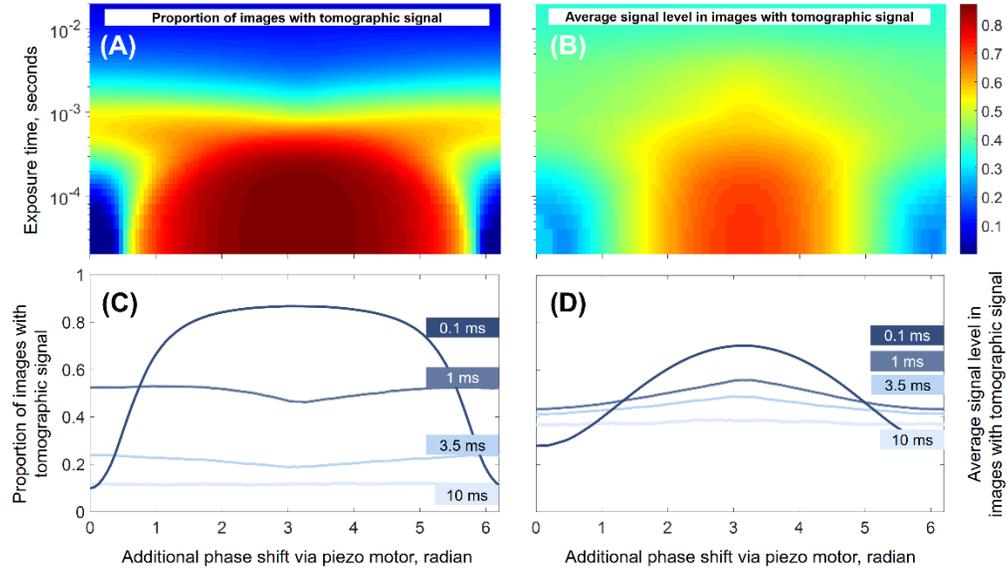

**Fig. 7.** (A) Proportion of tomographic images in presence of eye movements depending on the exposure time and additional piezo phase shift. By tomographic images we mean images with tomographic signal above 20% (arbitrary selected). Here the exposure time is the period, required to capture a single tomographic image (or the two camera frames). The 2D surface plot is calculated from Eq. 2 with experimental data from Fig. 3(A). (B) Normalized average signal level of tomographic images depending on the exposure time and additional phase shift. (C) Extracted slices from (A). (D) Extracted slices from (B). Horizontal scale is similar for all plots. Color scale is similar for (A) and (B).

Next, we tested the ability of TD-FF-OCT to acquire tomographic images without using a piezo motor. The exposure time (2 camera frames) was set to 3.5 ms (275 fps), 10 ms (100 fps) and 20 ms (50 fps). In all cases we were able to acquire images both with and without piezo motor modulation (Fig. 8). Stromal nuclei were visible in all images. However, in the case of long exposure (20 ms) the lateral movements created artifacts. More precisely, when the lateral movements are large, the incoherent light outside of the coherence volume is not completely removed during the consecutive image subtraction (see Eq. 1). Most of the incoherent light originates from the surface reflection from the tear film, therefore in the final image the contrast of the tomographic signal over background is reduced and we see a defocused view of the corneal surface. More details on the influence of motion are provided below in the sub-section 3.6.

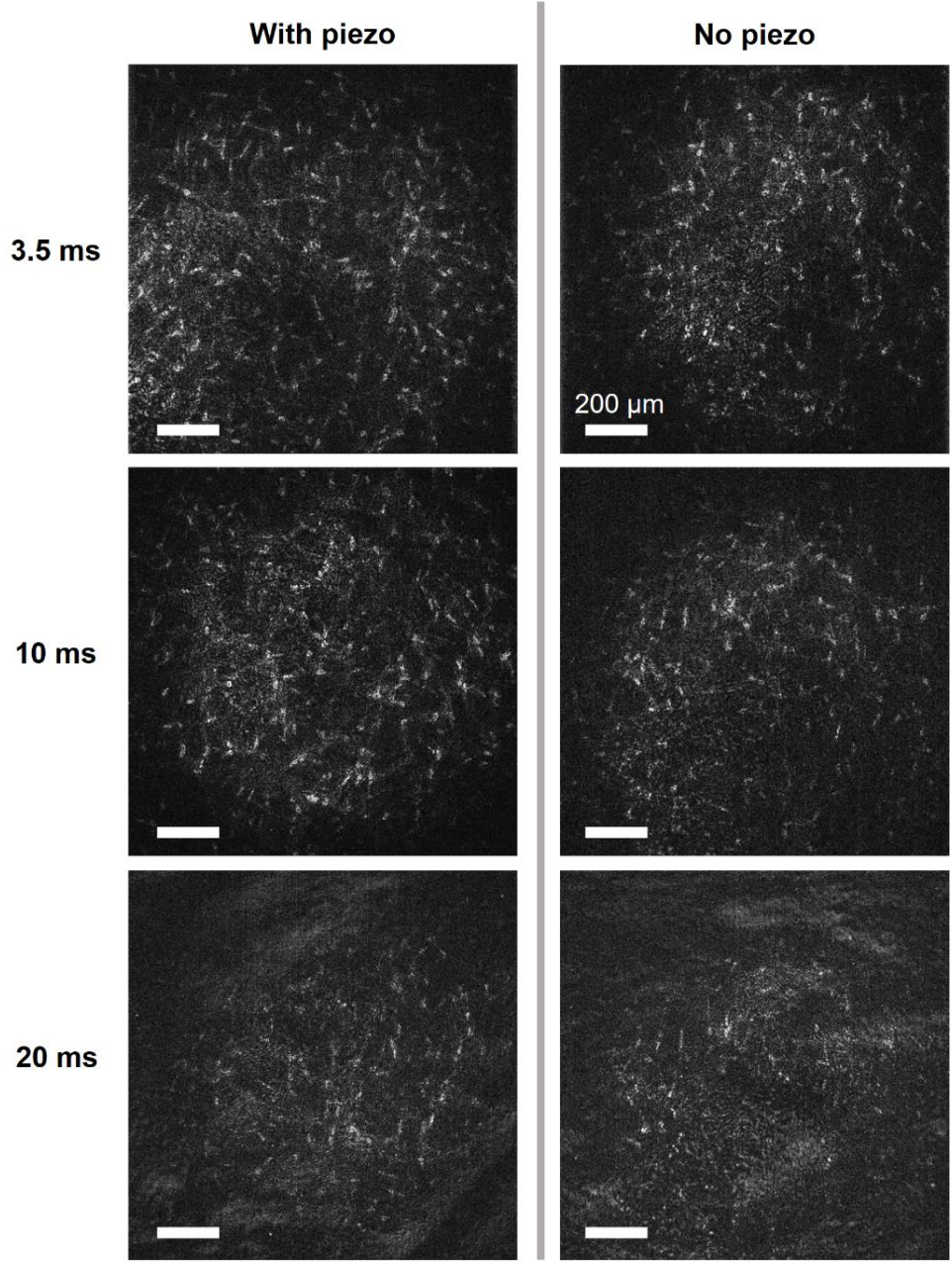

**Fig. 8.** Comparison of TD-FF-OCT images of human corneal stroma in vivo with and without piezo modulation at different exposure times. Stromal keratocytes are resolved. Exposure time here is the period required to capture two camera frames (single tomographic image). At long exposure time of 20 ms images contain artifacts related to the defocused view of the corneal surface. All scale bars are 200 μm.

*3.4 Comparison between healthy subjects*

The experiment and data analysis were repeated on another healthy subject, leading to similar results (Fig. 9). This further highlights the role of common physiological eye movements (such as heartbeat) in fast axial movements and phase modulation.

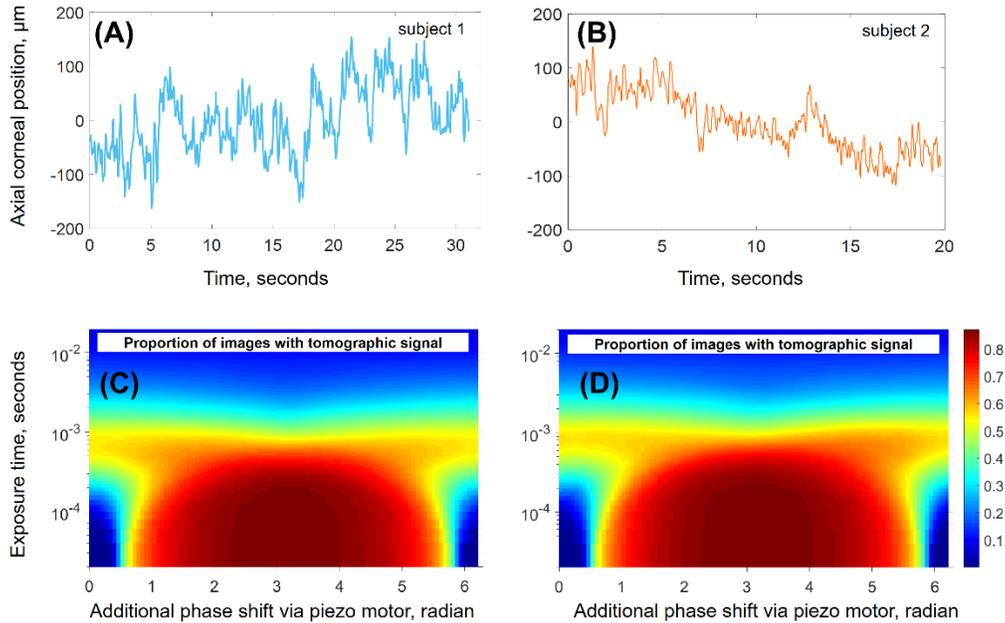

**Fig. 9.** (A), (B) Comparison of axial movements of in vivo human cornea from two healthy subjects. The source data underlying the plots are provided in **Data File 1**. (C), (D) Comparison of proportions of tomographic images depending on the exposure time and additional piezo phase shift for two healthy subjects. Color scale is similar for (A) and (B).

*3.5 Reconstruction of TD-FF-OCT signals from multiple images*

Previously we explored the TD-FF-OCT signals reconstructed from only two camera frames. However, thanks to the fast acquisition speed more images are typically available from the same coherence volume. These multiple images can be used to either increase the signal-to-noise ratio (SNR) by averaging or to remove the interference fringe artifacts (4 phase image retrieval scheme). Figs. 10 and 11 compare the effectiveness of different image averaging schemes in presence of eye movements and without piezo modulation: averaging of 40 tomographic images, standard deviation of 40 tomographic images and standard deviation of 40 direct camera frames. It should be noted that each image is being acquired with a different phase determined by the eye movements. All the schemes reduce the proportion of images having minimal (0%) and maximal (100%) signals, while increasing the proportion of middle level signals. The schemes differ in the most probable signal level. For averaging the majority of images have signal level of 10%, while for standard deviation of tomographic images and direct images this level is 20% and 30% respectively. Although the latter method produces large proportion of images with the high tomographic signal, it is not practical as the standard deviation of direct images amplifies not only the signals originating from the coherence volume, but also the motion of irregularities on the ocular surface (Fig. 11). More details on the influence of motion are provided below in the sub-section 3.6.

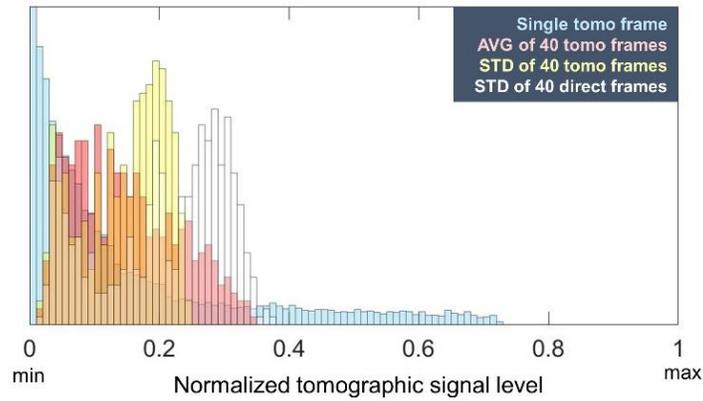

**Fig. 10.** Distribution of normalized tomographic signal level under different image averaging schemes. The histograms are calculated from Eq. 2 considering the experimental data in Fig. 3, while assuming a 3.5 ms exposure time, 850 nm light wavelength and no piezo modulation. Averaging (AVG) or taking the standard deviation (STD) reduces the proportion of images with minimal (0%) and maximal (100%) signal, while increasing the proportion of mediocre signals.

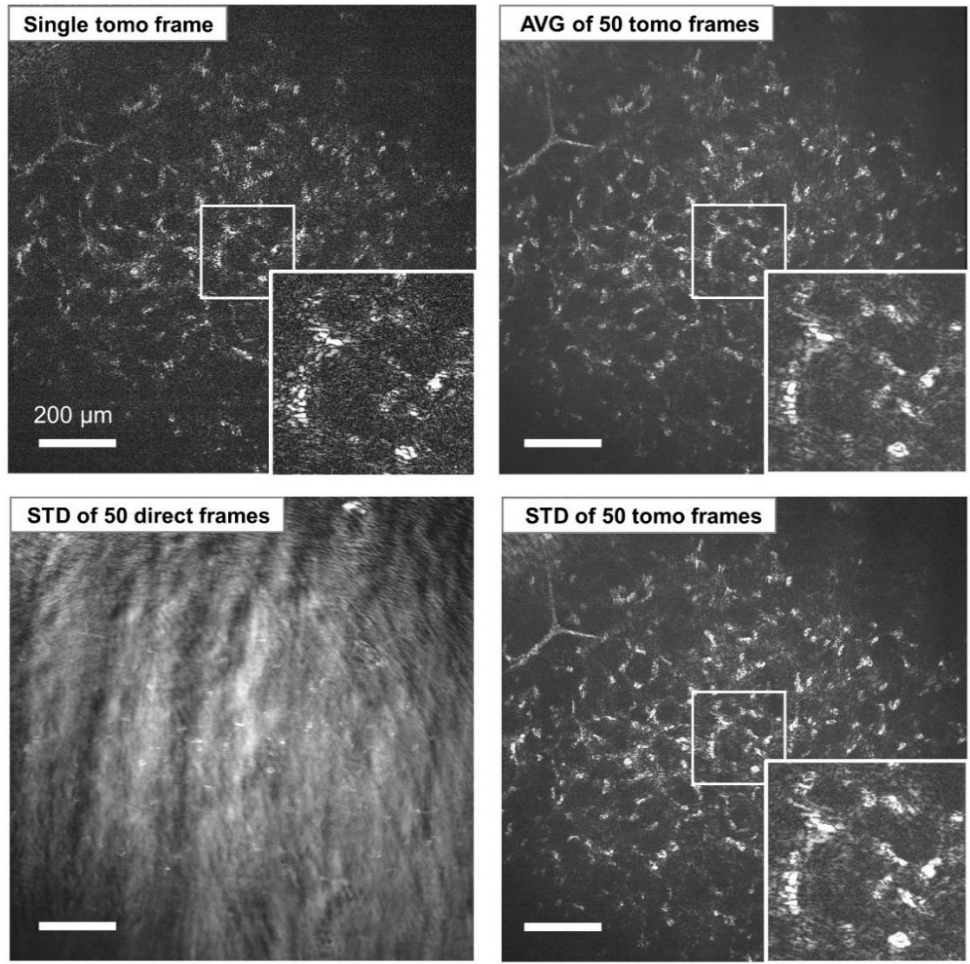

**Fig. 11.** Comparison of TD-FF-OCT images from in vivo human corneal stroma reconstructed from multiple frames via averaging or standard deviation. All scale bars are 200 μm.

Another question that is worth asking: can eye movements be used to acquire tomographic images under a four-phase modulation scheme (in comparison to the two-phase scheme tested before). The four-phase modulation scheme has an advantage, that it suppresses the pixel phase $\varphi(x, y)$ term (Eq. 1), which is responsible for the fringe artifacts, present in every two-phase tomographic image [30,31,34]. While this is of low importance in the irregular tissue interior, the artifacts are particularly visible in uniform reflective layers, such as the corneal endothelium. The typical four-phase modulation scheme reconstructs the tomographic image from the four camera frames having phases: 0, π/2, π, 3π/2. From the eye motion data (Fig. 3(A)) we calculated the 8% probability for the sum error in phases of four consecutive images to be below 3π/4 (less than π/4 error in each phase shift):

$$\left|(\psi_1-\psi_2)-\pi/2\right| + \left|(\psi_1-\psi_3)-\pi\right| + \left|(\psi_1-\psi_4)-3\pi/2\right| < 3\pi/4 \qquad (7)$$

Although this probability is low, we were able to acquire four phase images and reveal the endothelial cell mosaic in single TD-FF-OCT frames without the need for frame averaging (Fig. 12).

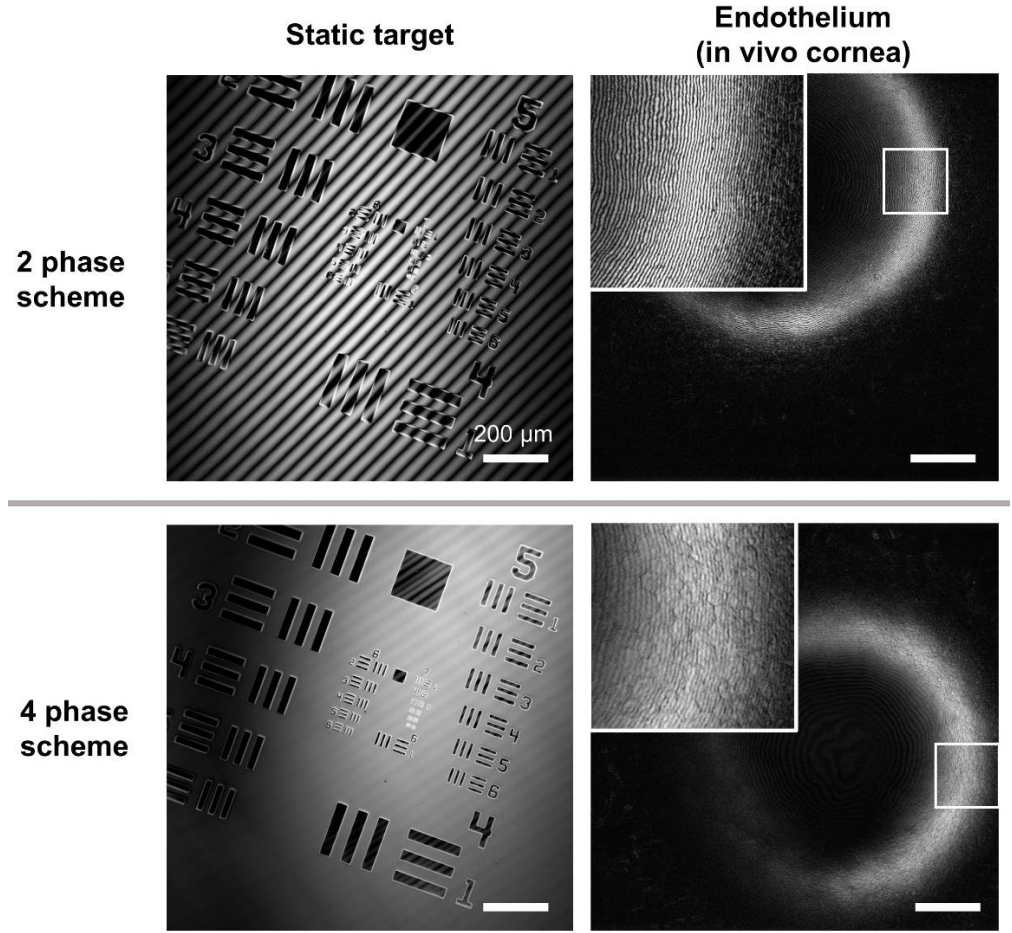

**Fig. 12.** Comparison of two- and four-phase modulation schemes. In vivo corneal images were acquired without piezo motor. The total exposure times were 3.5 ms for 2 phase and 7 ms for 4 phase images. Target images were acquired using the piezo modulation. The four-phase scheme suppresses the fringe artifacts that are typically present in uniform layers, such as the endothelium. All scale bars are 200 μm.

*3.6 Influence of lateral motion*

Lastly, we explored the influence of the lateral motion on TD-FF-OCT images. The two fastest sources of lateral motion are the saccades, rotating the eye at about 14°/s [35], and the movement of the tear film, spreading after the blink at about 4 mm/s [29].

First, we checked, if the saccadic lateral movement can shift the corneal axial position (leading to an axial optical phase shift). Assuming that the center of the corneal curvature does not match with the rotation axis of the eyeball, we theoretically estimated the negligible axial corneal shift on a millisecond timescale (~ 0.3 nm/ms). Although, saccades do not produce significant axial shift, they substantially shift the cornea laterally (~ 3 µm/ms). This shift is larger than the typical lateral resolution in TD-FF-OCT microscope, thus it affects the tomographic images. More precisely, when the shift between the consecutive camera images is large, the incoherent light outside of the coherence volume is not completely removed during the consecutive image subtraction (see Eq. 1). As most of the incoherent light originates from the air – tear film interface, the tomographic image reveals artifacts of structures from the corneal surface (Fig. 13). Saccadic motion shifts the structures within the coherence volume as well, which also suppresses the tomographic signal.

Following the blink, the tear film moves rapidly (~ 4 µm/ms), but independently from the underlying static cornea. Therefore, it produces the same incoherent surface artifacts as saccades, but does not significantly reduce the tomographic signal. One second after the blink the tear film velocity reduces below (~ 1 µm/ms) [29] and almost all lateral artifacts disappear.

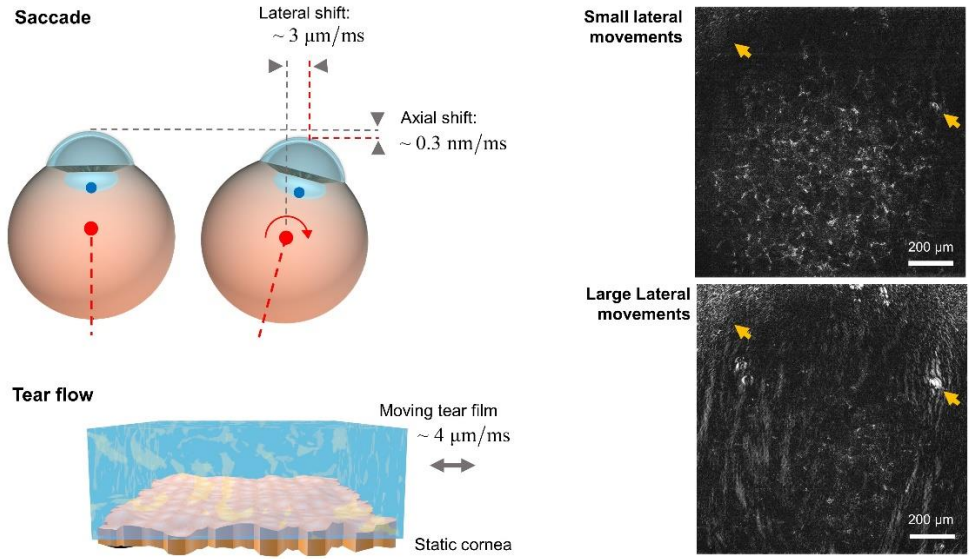

**Fig. 13.** Sources of lateral motion and their effects on TD-FF-OCT signal. On millisecond timescale saccades produce small axial and large lateral shifts. The large lateral shifts create the incoherent artifacts, more precisely the defocused view of the corneal surface (yellow arrows), and reduce the tomographic signal. Both images on the right were acquired from the same plane in human cornea in vivo. Tear flow produces similar surface artifacts during one second following the blink, while the tear velocity is high.

## 4. Discussion and Summary

In this article we have studied the axial movements of the human eye at different time scales. The acquired knowledge about the distribution of axial eye movements can potentially be useful for improving existing axial eye tracking and correction methods, such as described in [29,33,36–42]. During the exposure times below 0.1 ms the eye shifts by less than 100 nm, equivalent to an optical phase shift of $\pi/2$ for an 850 nm light wavelength. This confirms immunity of fast microsecond-exposure instruments to phase washout. During more than 1 ms the eye can move axially by more than 1 μm, leading to large optical phase shifts of π, 2π, 3π, etc. While being destructive for spectral-domain OCT methods, these large phase shifts can be useful in time-domain methods for path-length modulation within the interferometer, substituting the commonly used piezo motor. However, it should be noted that modulation with eye movements only (without piezo) limits the maximal achievable signal to 70% (compared to the theoretical 100% maximum with static sample and piezo modulation). This is due to the continuous linear nature of eye motion on a millisecond timescale. By adding the discontinuous step-like piezo modulation, 100% of the signal can be recovered.

The proportion of images having sufficient tomographic signal depends on the exposure time and amplitude of the piezo modulation. When the exposure time is relatively large above 1 ms, the phase modulation is dominated by the ocular movements and the probability of getting tomographic images is not affected by the piezo modulation. For example, the proportion is fixed at 50% (every 2$^{nd}$ image) for a 1 ms exposure. Below 1 ms, the speed of eye movements gradually becomes insufficient to provide the necessary π phase shift between the frames. In this case the phase modulation with piezo motor becomes essential for capturing the tomographic signals.

Experimentally, tomographic images of the in vivo human cornea could be revealed with the TD-FF-OCT instrument at 3.5 ms (275 fps), 10 ms (100 fps) and 20 ms (50 fps) exposure times (two camera frames) using only the phase shift, induced by the eye movements without piezo motor. This opens a path to a simplified ophthalmic TD-FF-OCT device, which uses a cheaper low-speed camera and avoids the piezo motor-camera synchronization chain. It should be noted that: 1) TD-FF-OCT needs to acquire at least two frames on the camera (at 100 fps) to reconstruct the tomographic image (at 50 fps), 2) the lateral eye movements set the lower bound for the feasible lowest camera speed.

A stack of multiple tomographic images originating from the same coherence volume can be processed to obtain a high SNR image even, if each image is being acquired with a different phase determined by the eye movements.

Finally, by using the natural eye movements it is possible to acquire not only two-phase, but also four-phase tomographic images. The latter have a benefit in that they suppress the unwanted interference fringe artifacts that are particularly visible on deterministic interfaces. The probability of getting the ideal phase shifts between the four consecutive images is negligibly low, nevertheless perfect matching is not required and visibility of interference fringe artifacts can be considerably reduced. This enables imaging of corneal endothelial cell mosaic using a single tomographic frame.

**Funding**

This work was supported by the HELMHOLTZ synergy grant funded by the European Research Council (ERC) (610110), a CNRS pre-maturation grant, Region Ile-De-France fund SESAME 4D-EYE (EX047007), French state fund CARNOT VOIR ET ENTENDRE (x16-CARN 0029-01), French state fund IHU FOReSIGHT (ANR-18-IAHU-0001), PSL pre-maturation grant under the French Government program "Investissements d'Avenir" (ANR-10-IDEX-0001-02 PSL) and proof of concept (POC) grant funded by the European Research Council (ERC) (957546).

**Acknowledgments**

We would like to acknowledge the advisory support of the Quinze-Vingts National Ophthalmology Hospital.

**Disclosures**

VM: (P), PX: (P), MF: (P), ACB: (P).

## Data Availability

Data underlying the results presented in this paper are available in Data File 1.